\documentclass[twocolumn,amsmath,aps]{revtex4}
\usepackage{graphicx}

\newcommand{\nl}{\nonumber \\}

\newcommand{\be}{\begin{equation}}
\newcommand{\ee}{\end{equation}}
\newcommand{\bea}{\begin{eqnarray}}
\newcommand{\eea}{\end{eqnarray}}

\newcommand{\Eq}[1]{Eq.\,(\ref{#1})}

\newcommand{\la}{\langle}
\newcommand{\ra}{\rangle}
\newcommand{\dg}{\dagger}

\newcommand{\ti}{\tilde}
\newcommand{\mb}{\mbox}

\begin{document}
\draft

\title{Quantum trajectory analysis for electrical detection of single-electron
spin resonance}

\author{Jinshuang Jin, Jianhong Guo, Junyan Luo and Xin-Qi Li}
\address{ State Key Laboratory for Superlattices and Microstructures,
                  Institute of Semiconductors,Chinese Academy of Sciences,
                  P.O. Box 912, Beijing 100083, China}
\author{YiJing Yan}
\address{Department of Chemistry, Hong Kong University of Science and
         Technology, Kowloon, Hong Kong}

\date{\today}

\begin{abstract}
A Monte Carlo simulation on the basis of quantum trajectory approach
is carried out for the measurement dynamics of a single electron
spin resonance. The measured electron, which is confined in either
a quantum dot or a defect trap, is tunnel-coupled to a side
reservoir and continuously monitored by a mesoscopic detector. The
simulation not only recovers the observed telegraphic signal of detector
current, but also predicts unique features in the output power spectrum
which are associated with electron dynamics in different regimes.
\end{abstract}

\vspace{3ex}
\maketitle

In recent years, largely being stimulated by the interest of
solid-state quantum computation, measurement of single electron spin
has become an intensive research subject.
The main difficulty of spin measurement lies in the inherent weakness of
magnetic interaction, making its direct detection quite challenging.
The current state-of-the-art techniques include, for instances,
scanning tunneling microscopy \cite{Ber01},
magnetic resonance force microscopy (MRFM) \cite{Ber03},
sub-micron Hall magnetometer \cite{Jin05},
and spin-to-charge conversion \cite{Los03}, etc.
Experimentally, detection of single electron spin has been
illustrated by optical means \cite{Gam01,Jel03},
MRFM technique \cite{Rug04}, and electrical methods \cite{Kou04,Mar04}.
In particular, the idea of electrical approach is based on
the spin-to-charge conversion, which maps the spin-states onto
detectable charge states.
In Refs.\ \onlinecite{Kou04} and \onlinecite{Mar04}
single shot read-out of the charge state is demonstrated in terms of
single realization of continuous measurement.
In this work, we present a Monte Carlo simulation for the dynamics
of continuous electrical measurement, and calculate the output noise
spectrum. This study provides a fundamental description for the
stochastic nature of state evolution, which stems from the interplay
of the manipulation of coherent field and the (simultaneous)
influence of incoherent reservoir. Also, the predicted unique
features in the noise spectrum manifest distinct electron dynamics
in different regimes.

\begin{figure}\label{Fig1}
\includegraphics*[scale=0.4,angle=0.]{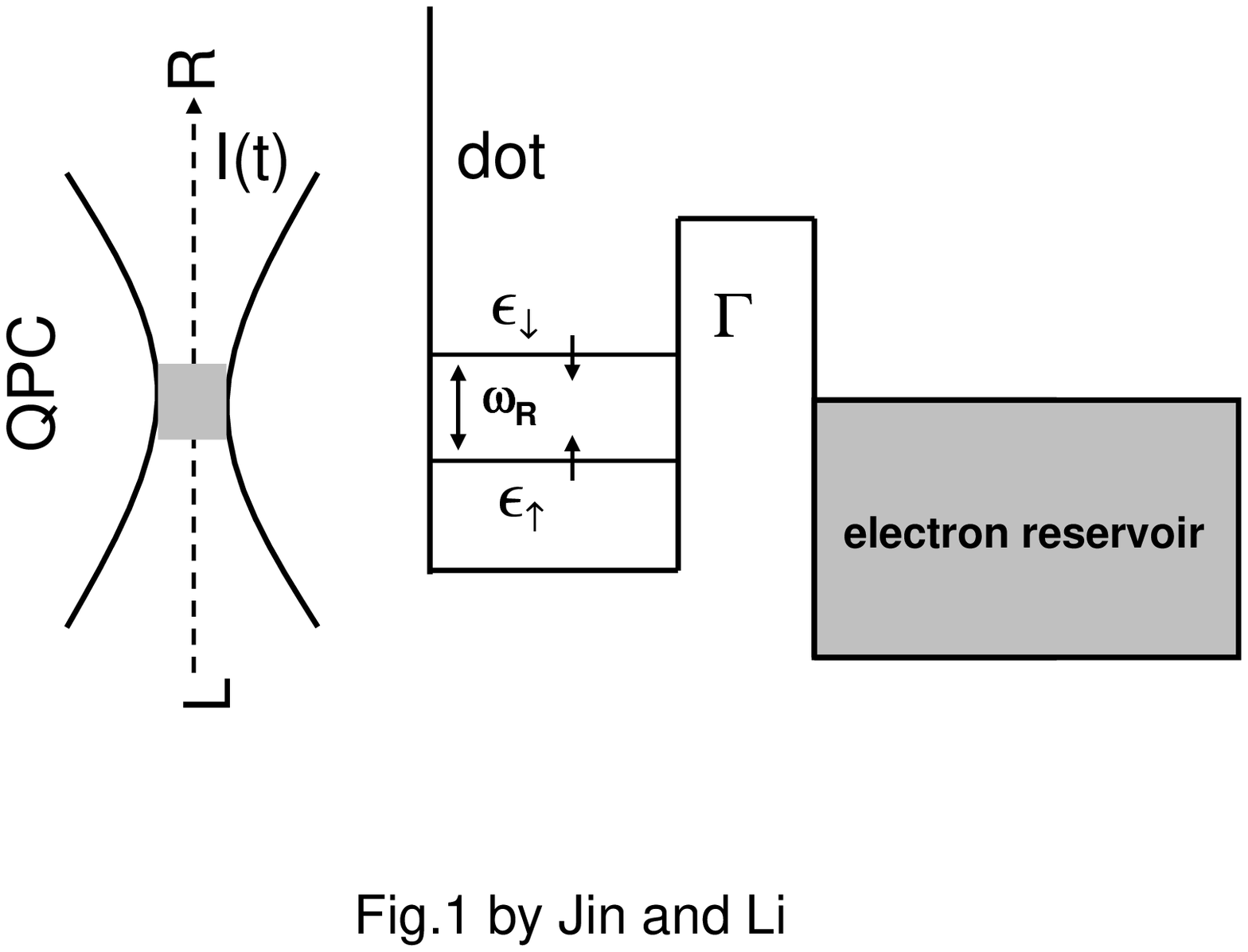}
\caption{Schematic diagram of a quantum dot being tunnel-coupled to
a side reservoir and measured by a quantum point contact (QPC). The
single-electron level in the dot is split by an external magnetic
field and the two spin states are coupled by an electron spin
resonance (ESR) magnetic field. The tunneling of single electron
into and out of the dot would alter the transport current through
the QPC detector.}
\end{figure}

In close connection with the experiments \cite{Kou04,Mar04},
we consider a model system of quantum dot in the Coulomb blockage regime,
which is tunnel-coupled to a side electronic reservoir and
continuously monitored by an electrical detector, say,
the quantum point contact (QPC),
as schematically shown in Fig.\ 1.
The single-electron level in the dot is split
by an external magnetic field $B_0$,
i.e., with the Zeeman energy of
$\epsilon_{\downarrow}-\epsilon_{\uparrow}=g\mu_B B_0\equiv\Delta$,
where $g$ is the Land$\acute{e}$-$g$ factor of electron spin
and $\mu_B$ is the Bohr magneton.
The spin-up and spin-down states are coupled by an oscillating
magnetic field $B_{\rm ESR}=B_1 \cos(\omega t)$,
applied perpendicularly to the field $B_0$, which
leads to the well known electron spin resonance (ESR)
with Rabi frequency $w_R=g\mu_B B_1$.
In the Coulomb blockade regime, the dot is occupied by at most one electron.
Thus, the dot electron Hamiltonian in the presence of magnetic fields
reads $H_{\rm dot}=\frac{\Delta}{2}(d_{\downarrow}^{\dg}d_{\downarrow}
-d_{\uparrow}^{\dg} d_{\uparrow})
+\frac{\omega_R}{2}(d_{\downarrow}^{\dg}d_{\uparrow}e^{-i\omega t}
+d_{\uparrow}^{\dg}d_{\downarrow}e^{i\omega t})$,
in which the first part describes the Zeeman splitting,
and the second one represents the electron spin resonance.
Following the QPC measurement model proposed by Gurvitz \cite{Gur97},
the entire measurement setup shown in Fig.\ 1 is described by the Hamiltonian
$H =  H_{0}+H_{T}+H_{\rm int}$,
with
\begin{subequations} \label{H1}
\begin{eqnarray}
H_0&=&H_{\rm dot}+H_{\rm res}+H_{\rm PC}
\\
H_T &=& \sum_{q\sigma}[t_q c_{q\sigma} ^{\dg}d_{\sigma}+H.c.]
\\
 H_{\rm int} &=& \sum_{k,q} (T_{qk}+\chi_{qk}n_0)
         a^{\dg}_{Lk} a_{Rq}  + \mb{H.c.}.
\end{eqnarray}
\end{subequations}
Here the Hamiltonian
$H_{\rm res}=\sum_{q\sigma}\epsilon_q c_{q\sigma} ^{\dg}c_{q\sigma}$
stands for the side electronic reservoir,
and $H_T$ describes its tunnel-coupling to the quantum dot.
$\sigma=\uparrow,\downarrow$ denote the two possible orientations
of electron spin.
For the QPC detector,
$ H_{\rm PC} = \sum_k \epsilon^L_k a^{\dg}_{Lk} a_{Lk}
+\sum_q\epsilon^R_q a^{\dg}_{Rq} a_{Rq}$
stands for the two (left and right) reservoirs,
and $H_{\rm int}$ describes tunneling between them.
Notice that the tunneling amplitude explicitly depends on the
electron occupation of the quantum dot, which is modelled by
$T_{qk}+\chi_{qk}n_0$,
where $n_0\equiv 1-\sum_{\sigma}d^{\dg}_{\sigma}d_{\sigma}$
is the vacancy operator of the quantum dot.

{\it Reduced description for the measurement process}.
--- A significant aspect of quantum measurement is to investigate the back-action
of measurement device on the measured system, which results in the
dephasing and relaxation.
Technically, this can be realized by tracing out the microscopic
degrees of freedom of the measurement device.
Physically, this reduced description corresponds to ensemble results, i.e.,
all the measurement records are averaged.
For present setup,
the side reservoir and the detector can be treated as two independent
environments acting on the (measured) quantum dot.
As a model description,
we assume the tunneling amplitude $T_{kq}$ to be real and independent
of the reservoir-state ``$kq$", i.e., $H_{\rm int}=QF$,
where $Q=T+\chi n_0$
and $F=\sum_{kq}(a_{Lk} ^{\dg} a_{Rq} + {\rm H.c.})$.
For the coupling between the dot and the side reservoir,
we re-express the coupling Hamiltonian as
$H_T=\sum_{\sigma}(d_{\sigma}f_{\sigma} ^{\dg} + {\rm H.c.})$,
where $f_{\sigma}^{\dg}\equiv\sum_q t_q c_{q\sigma}^{\dg}$.
By treating $H_{\rm T}$ and $H_{\rm int}$ as perturbation
and on the basis of the second-order cummulant expansion,
an equation for the reduced density matrix
can be derived as \cite{Li04, Li05}
\bea \label{UME}
\dot{\rho} =
&-&i {\cal L}_0\rho  -\frac{1}{2}[Q,\ti{Q}\rho-\rho\ti{Q}^{+}]   \nl
&-&\frac{1}{2}\sum_{\sigma}\left\{ [d^{\dg}_{\sigma},D^{(-)}_{\sigma}\rho-
\rho D^{(+)}_{\sigma}]+H.c.\right\} .
\eea
Here, the first term ${\cal L}_0\rho \equiv [H_{\rm dot} ,\rho]$
describes the coherent manipulation of the dot electron (by the
ESR magnetic field), while the second and third terms stem from
the back-action of the detector and the random tunneling caused by
the side reservoir.
In the back-action term, the operator $\ti Q=\ti Q^{(+)}+\ti Q^{(-)}$,
and $\ti{Q}^{(\pm)}=C^{(\pm)}( {\cal L}_0) Q $,
where the superoperators $C^{(\pm)}({\cal L}_0)$ are the Fourier transform
of the correlation functions $C^{(+)} (t)=\la F(t)F(0) \ra$
and $C^{(-)} (t)=\la F(0)F(t) \ra$.
Under the wide-band approximation, $C^{(\pm)}({\cal L}_0)$
can be explicitly carried out as \cite{Li04}:
$C^{({\pm})} ({\cal L}_0)
 =\eta\left[\frac{x}{1-e^{-\beta x}}\right]_{x=-({\cal L}_0 \pm e V)}$.
Here, $\eta=2\pi g_l g_r$, with $g_l$ and $g_r$ the energy-independent density
of states for the two reservoirs of the QPC detector.
$\beta=1/(k_B T)$ is the inverse temperature,
and $eV=\mu_L-\mu_R$ is the applied voltage across the detector.
Similarly, for the side reservoir, we obtain \cite{Li05}:
$D^{(\pm)}_{\sigma}=\Gamma_{\sigma}n^{(\pm)}_{\sigma}
(\epsilon_{\sigma})d_{\sigma}$.
Here $n^{(+)}_{\sigma}(\epsilon_{\sigma})=n_{\sigma}(\epsilon_{\sigma})$
and $n^{(-)}_{\sigma}(\epsilon_{\sigma})=1-n_{\sigma}(\epsilon_{\sigma})$,
with $n_{\sigma}$ the Fermi distribution function.
$\Gamma_{\sigma}=2\pi g_c |t_q|^2$,
in which $g_c$ is the density of states of the side reservoir electrons
at the energy $\epsilon_{\sigma}$.

At the limit of large bias voltage \cite{Gur97, Mil01}, \Eq{UME}
can be further simplified to the Lindblad-type master equation
\bea\label{rho}
\dot{\rho}=&-&i {\cal L}_0\rho+\Gamma_{\downarrow}
{\cal D}[S_1]\rho+\Gamma_{\uparrow}{\cal D}[S_2]\rho \nl &+&{\cal
D}[{\cal T}+{\cal X} n_0]\rho\equiv{\cal L} \rho(t).
\eea
Here we have also restricted our study to zero temperature. The
superoperator ${\cal D}$ is defined as ${\cal D}[r]\rho={\cal
J}[r]\rho-{\cal A}[r]\rho$, where ${\cal J}[r]\rho\equiv r\rho
r^{\dg}$, and ${\cal A}[r]\rho\equiv \frac{1}{2}(r^{\dg}r\rho+\rho
r^{\dg}r)$. The jump operators $S_1\equiv|0\ra\la\downarrow|$ and
$S_2\equiv|\uparrow\ra\la 0|$ describe the tunneling between the
dot and the side reservoir.
The {\it homodyne-type} jump operator ${\cal D}[{\cal T}+{\cal X} n_0]$,
which describes the back-action of the detector, is associated with the tunneling
amplitudes through the QPC with and without an electron in the
quantum dot, by the corresponding tunneling rates
$|{\cal T}|^2=\eta |T|^2 eV\equiv D$, and $|{\cal T}+{\cal X}|^2=\eta
|T+\chi|^2 eV\equiv D'$.
Note that, differing from the setup of a charge qubit measured by
a QPC \cite{Gur97}, here the QPC's back-action has minor effect on
the measured-electron dynamics. The reason is that, in the
occupation state basis of the quantum dot, ${\cal D}[{\cal
T}+{\cal X} n_0]\rho$ only contributes to the off-diagonal
elements, which are decoupled from the diagonal ones. That is, the
back-action only causes dephasing between the ``occupied" and
``empty" dot states. But no phase coherence is introduced
initially, and neither will it be generated in the later on
evolution. Thus the QPC has no influence on the electron dynamics
under study \cite{note-1}.

Straightforwardly, the (ensemble) measurement current through the detector
can be calculated via
\bea\label{I1}
I=I_0 \rho_{00}+I_1 \rho_{11},
\eea
where $I_0=eD'$ and $I_1=eD$ are, respectively,
the currents through the QPC in the absence and presence of
an electron in the quantum dot. Note that \Eq{I1} is precisely
the result presented in Ref.\ \onlinecite{Mar03}.
In the following we will see that it is indeed an ensemble average of the
continuously measured currents from a large number of individual realizations.

{\it Quantum trajectory description}.
--- The above reduced density matrix description
corresponds to average over all the measurement records
of many individual realizations.
However, in most cases the original readout data of
continuous measurement of a single quantum system is
the record from a single measurement realization.
For instance, the experimental result of Ref.\ \onlinecite{Mar04}
reveals random telegraph signal (RTS), with detector current jumping
between two discrete values stochastically.

Proper description for this single measurement realization
is the quantum trajectory theory developed in quantum optics
\cite{Scu97,Wis94},
which describes the system evolution conditioned on the information
continuously acquired by the detector.
The central ingredient of this formalism is
the conditional master equation (CME), which is
an unravelling of the unconditional master equation \Eq{rho}.

For the tunneling events between the quantum dot and the side reservoir,
two stochastic point variables $dN_1(t)$ and $dN_2(t)$
(with values either 0 or 1) are employed to denote,
respectively, the electron numbers that tunnel
from the dot to the side reservoir and vice versa,
during the infinitesimal time interval $dt$.
While for the QPC, at the limit of zero temperature and large bias voltage,
the electron only tunnels from the source (left reservoir)
to the drain (right reservoir), which is denoted by an alternative
stochastic variable $dN_c(t)$.
Following Ref.\ \onlinecite{Mil01}, we obtain
\bea\label{CME1}
d\rho_c(t)
&=&  dt \left\{-i{\cal L}_0-{\cal A}[{\cal T}+{\cal X} n_0]\right.\nl
 &&\left.-\Gamma_{\downarrow}{\cal A}[S_1]-\Gamma_{\uparrow}{\cal A}[S_2] \right.\nl
 && \left.+ {\cal P}_{c}(t)+ {\cal P}_{1}(t) + {\cal P}_{2}(t) \right\} \rho_c(t)\nl
 &&+dN_1(t)\left [\frac{\Gamma_{\downarrow}{\cal J}[S_1 ]}{{\cal P}_{1}(t)}-1\right ]
 \rho_c(t) \nl
 &&+dN_2(t)\left [\frac{\Gamma_{\uparrow}{\cal J}[S_2 ]}{{\cal P}_{2}(t)}-1\right ]
 \rho_c(t) \nl
 &&+dN_c(t)\left [\frac{{\cal J}[{\cal T}+{\cal X} n_0]}{{\cal P}_{c}(t)}-1\right ]
 \rho_c(t),
\eea
in which the stochastic variables $dN_i(t)~(i=c,1,2)$ have the ensemble property
$E[dN_i(t)]={\cal P}_{i}(t) dt={\rm Tr} \{{\cal J}[r_i]\rho_c(t)\}dt$, with
$r_c={\cal T}+{\cal X} n_0$, $r_1=\sqrt{\Gamma_{\downarrow}}S_1$
and $r_2=\sqrt{\Gamma_{\uparrow}}S_2$, respectively.
Here $E[\cdots]$ denotes an ensemble average of a
large number of stochastic processes.

Within the quantum trajectory approach, the current through the QPC is given by
\begin{equation}\label{It1}
I(t) = e\, {dN_c(t)}/{dt}.
\end{equation}
Here we show that its ensemble average recovers the current given by \Eq{I1}.
With the help of the CME \Eq{CME1}, the stationary-state current
can be evaluated as
\bea\label{Is}
I_{\infty}
&=&e E[{dN_c(t)}/{dt}]_{t\rightarrow\infty}  \nl
&=&e{\rm Tr}\left\{{\cal J}[{\cal T}+{\cal X}n_0]\rho(\infty)\right\} \nl
&=&e\left[D+(D'-D)\frac{\omega ^2 _R}{\Gamma^2+3\omega ^2 _R+4\delta^2}\right],
\eea
where $\rho(\infty)$ is the ensemble stationary state calculated by \Eq{rho}. Here,
we have assumed $\Gamma_{\uparrow}=\Gamma_{\downarrow}\equiv\Gamma$.
We notice that \Eq{Is} is precisely the same as derived in
Ref.\ \onlinecite{Mar03} based on \Eq{I1}.

For the typical mesoscopic detector QPC, the condition $D\gg |D'-D|$
implies a large background current.
Accordingly, it is desirable to convert the above point process description
into a diffusive one \cite{Mil01}.
First, the terms in \Eq{CME1},
$\{-{\cal A}[{\cal T}+{\cal X}n_0]+{\cal P}_c(t)\}\rho_c(t)dt+
dN_c(t)\left [\frac{{\cal J}[{\cal T}+{\cal X} n_0]}{{\cal P}_{c}(t)}-1\right ]
 \rho_c(t)$,
are replaced by
 ${\cal D}[{\cal T}+{\cal X} n_0]\rho_c(t)dt
+{\cal X}[n_0\rho_c(t)+\rho_c(t)n_0-2\langle n_0\rangle\rho_c(t)]\xi(t)dt$,
where $\xi(t)$ is a Gaussian white noise characterized by $E[\xi(t)]=0$ and
$E[\xi(t')\xi(t)]=\delta(t-t')$.
Then, the detector current $I(t)$ of \Eq {It1} in the diffusive regime becomes
\bea\label{It2}
I(t)-\overline{I}=\Delta I [\rho^c_{00}(t)-\rho^c_{11}(t)]/2+\sqrt{S_0/2}\xi(t),
\eea
where $\overline{I}=(I_0+I_1)/2$ , $\Delta I=I_0-I_1$, and
the background shot noise $S_0=2eI_{\infty}$.
Here we introduced the notation
$\rho^c_{ij}(t)\equiv\la i|\rho_c|j\ra$, with ``$(i,j)=(0,1)$"
characterizing the number of electron occupied in the dot.

\begin{figure}\label{Fig2}
\includegraphics*[scale=0.45,angle=0.]{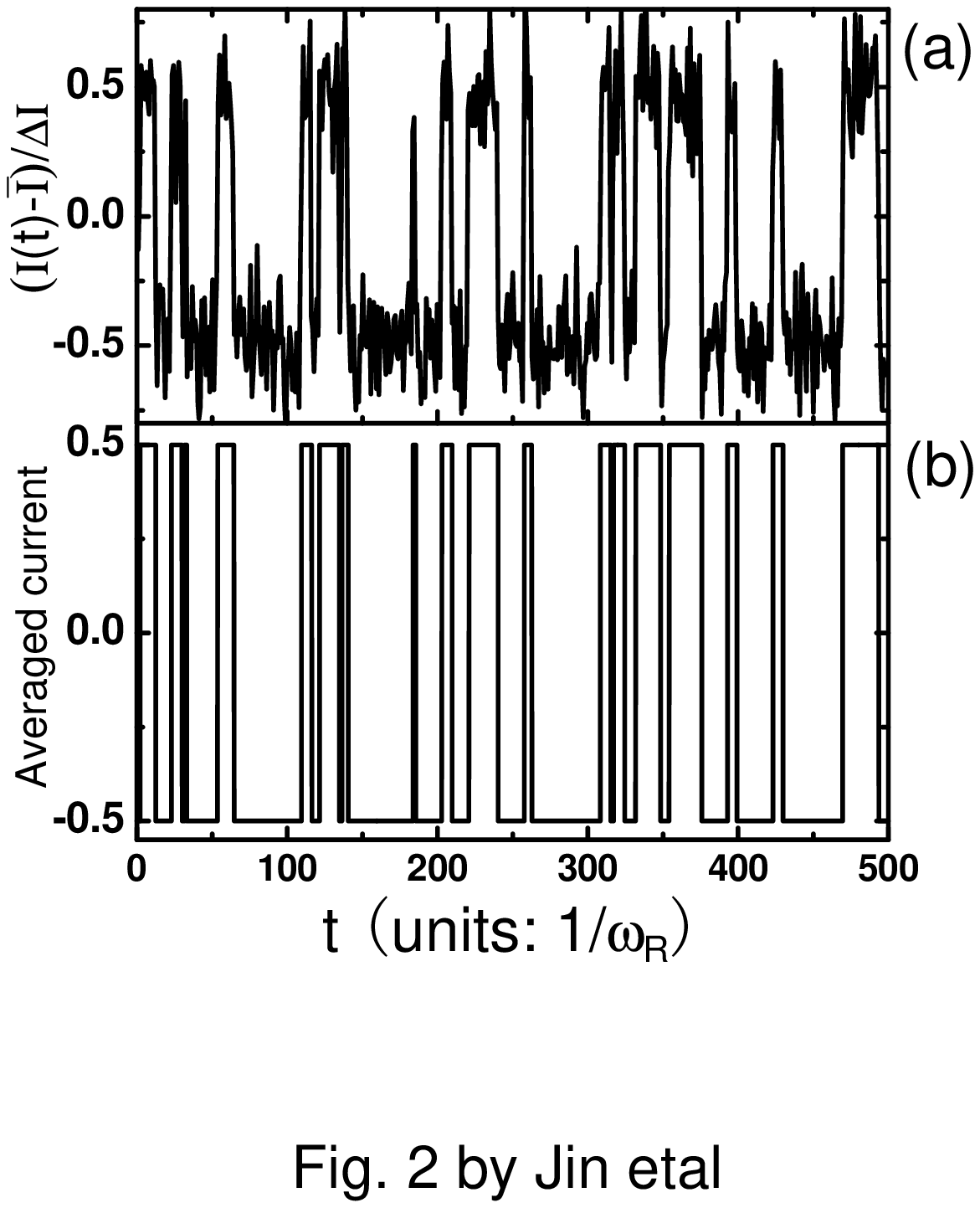}
\caption{ Detector current from a single particular realization of
quantum measurement, where the feature of random telegraph signal
(RTS) which jumps between two discrete current values is evident. In
the simulation, the resonance condition (i.e.
$\delta=\Delta-\omega=0$) and weak coupling between the dot and the
side reservoir (by setting $\Gamma=0.1\omega_R$) are assumed. The
simulated random current $I(t)$ in (a) is filtered by using a
rectangular window with width  $\tau=0.9/\omega_R$, and in (b) is
further averaged over the detector's background current
fluctuations. }
\end{figure}


From \Eq{It2} we find that the detector current $I(t)$ basically follows
the occupation probabilities
$\rho^c_{ii}(t)$ and contains a noise term $\xi(t)$ that
purely arises from the intrinsic fluctuations in the detector, being
independent of the measured system.
Particularly, in the Markov limit, the detector noise subjects to
Gaussian distribution as characterized above.
For noise strength weaker than the signal, single particular realization
of measurement will be evident.
Since the fluctuation grows when $I(t)$ is examined at smaller time
scales, proper averaging over time (``low-pass filtering") is
necessary \cite{Kor99}. Based on \Eq{It2}, we thus average $I(t)$
by using a rectangular filter window with time width $\tau$,
$I(t)=\tau ^{-1}\int _{t-\tau}^{t} I(t')dt'$.
The width of the filtering window should be larger than the
average time interval of two successive electron tunneling events through the
QPC junction, and smaller than the average occupation/empty time of the
quantum dot. This makes the output current not too noisy
in order to extract the information of the measured system.
The numerical result is shown in Fig.\ 2 (a), where the width of the filtering
window is chosen as $\tau=0.9/\omega_R$.
As a manifestation of the telegraph signal,
the current stochastically switches between two discrete values.
The low and high current plateaus correspond to
the quantum dot with and without an extra electron in it.
The average widths of the plateaus are determined by the
average occupation probabilities $\rho^c_{ii}(t)$.
Also, the detector shot noise would cause current fluctuations
around the plateaus.
If the detector's background current fluctuations are further averaged,
the telegraph signal will be more evident as shown  in Fig.\ 2 (b), which
agrees well with the experimental observation reported
in Ref.\ \onlinecite{Mar04}, in the regime $\Gamma\ll\omega_R$.

\begin{figure}\label{Fig3}
\includegraphics*[scale=0.4,angle=0.]{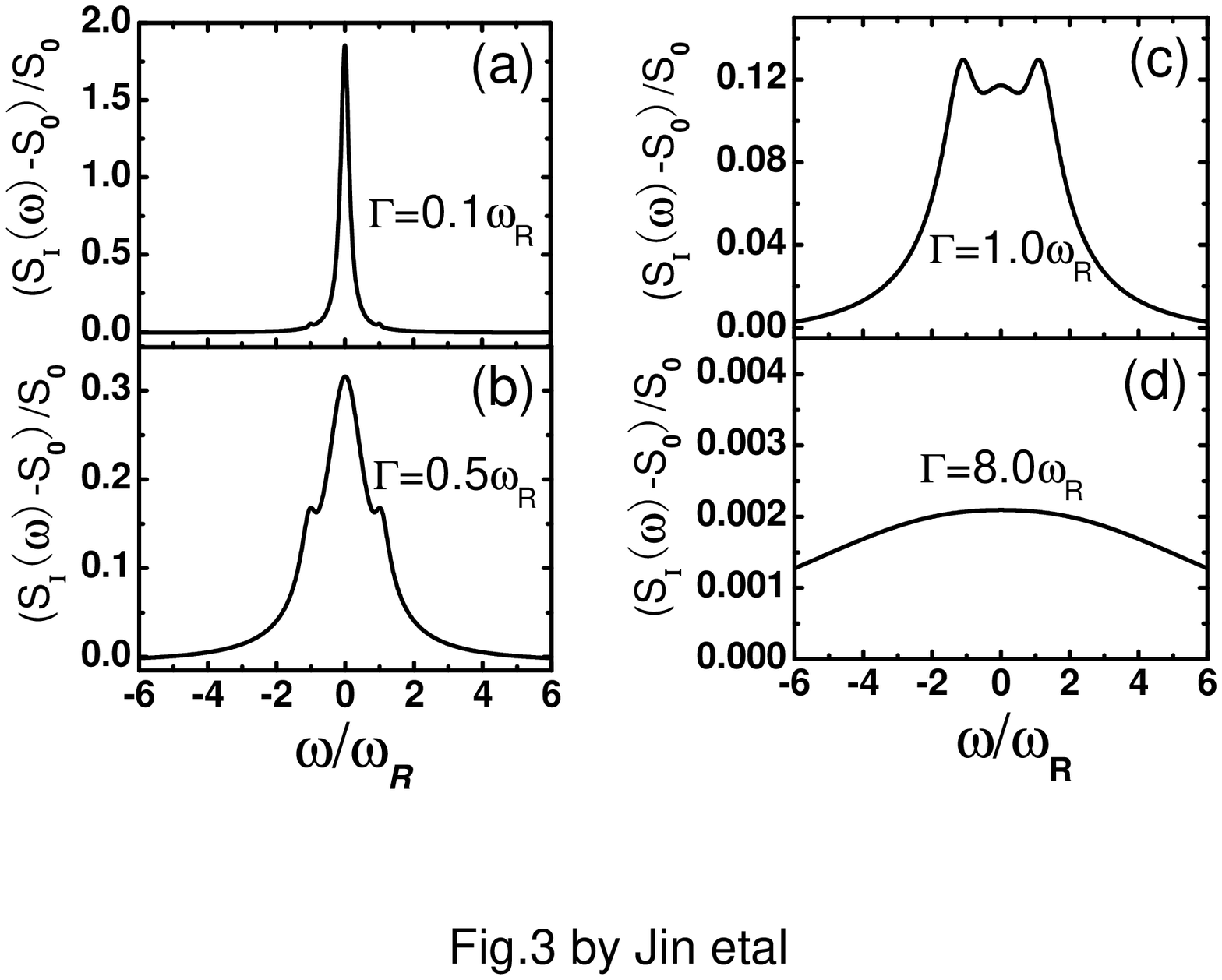}
\caption{Stationary noise power spectrum of the detector current.
(a)-(d) correspond to different coupling strengths $\Gamma$ between
the quantum dot and the side reservoir. }
\end{figure}

As we have seen in Fig.\ 2,
the single realization of the output current largely reflects
the instantaneous occupation of the quantum dot,
which is determined by the interplay of the coherent ESR driving
and the incoherent tunneling. Specifically, we can easily imagine:
(i) in the weak tunneling regime, the dot is either occupied or unoccupied
for a long time on average; (ii) in the intermediate tunneling regime,
the dot is occupied for longer time than unoccupied; and (iii) in the strong
tunneling regime, the dot is largely occupied.
Accordingly, the output current would follow these occupations and exhibit the
corresponding behaviors.
Rather than showing the entire landscape of the dot-electron dynamics
by the output current, in the following we show that it can be elegantly manifested
in the output power spectrum.

{\it Noise spectrum}.---
Following Refs.\,\onlinecite{Wis93} and \onlinecite{Goa01},
as deriving the ensemble-averaged current \Eq{Is}, starting with the current
formula \Eq{It1} which is defined in the ``point process" regime can also
lead to the current correlation function in the stationary state as
\bea\label{Gt}
G(\tau)&=&\left\{E[I(t+\tau)I(t)]-E[I(t+\tau)]
E[I(t)]\right\}_{t\rightarrow\infty}\nl
&=&eI_{\infty}\delta(\tau)+e^2(D'-D)^2\nl
&&\times\left\{{\rm Tr}\left[n_0 e^{{\cal L}\tau}{\cal J}[n_0]\rho_{\infty}\right]
-{\rm Tr}[n_0\rho_{\infty}]^2\right\}.
\eea
It should be noted that although this expression was derived in the
``point process" regime, it is still applicable in the diffusive regime,
since it is an ensemble averaged result.
By noting $G(-\tau)=G(\tau)$, the stationary noise power spectrum
can be computed by
\bea\label{Sw}
S(w)=2\int^{\infty}_{-\infty}d\tau G(\tau)e^{-iw\tau}.
\eea
Numerical results of the noise spectrum are shown in Fig.\ 3,
where the dot-electron dynamics under the interplay of the coherent ESR driving
and the incoherent tunneling is clearly manifested.
In the weak tunneling regime (i.e. $\Gamma\ll\omega_R$),
compared to the electron spin-up and spin-down flipping time in the dot,
the electron tunnels into and out of the dot slowly.
As a consequence, the dot occupation only changes occasionally with long-time
separation, which results in the dominant sharp Lorentzian peak around
the zero frequency, see Fig.\ 3(a).
This feature indicates nothing but the telegraph noise.
With the increase of $\Gamma$,
the in-dot coherent flipping will eventually play important role on
the electron's dynamics, which leads to the gradual suppression
of the central sharp peak (i.e. the telegraph noise spectrum),
and makes the coherent peak around the Rabi frequency more evident,
as shown by Fig.\ 3(b) and (c).
Finally, as $\Gamma\gg\omega_R$, the incoherent tunneling between the dot
and the side reservoir takes place so fast that the dot is almost occupied
for all time,
which greatly reduces the {\it excess} noise as shown in Fig.\ 3(d).
The considerably flattened peak shown there simply reflects the
sudden change of the dot occupation, which resembles the
white noise process.

\begin{figure}\label{Fig4}
\includegraphics*[scale=0.4,angle=0.]{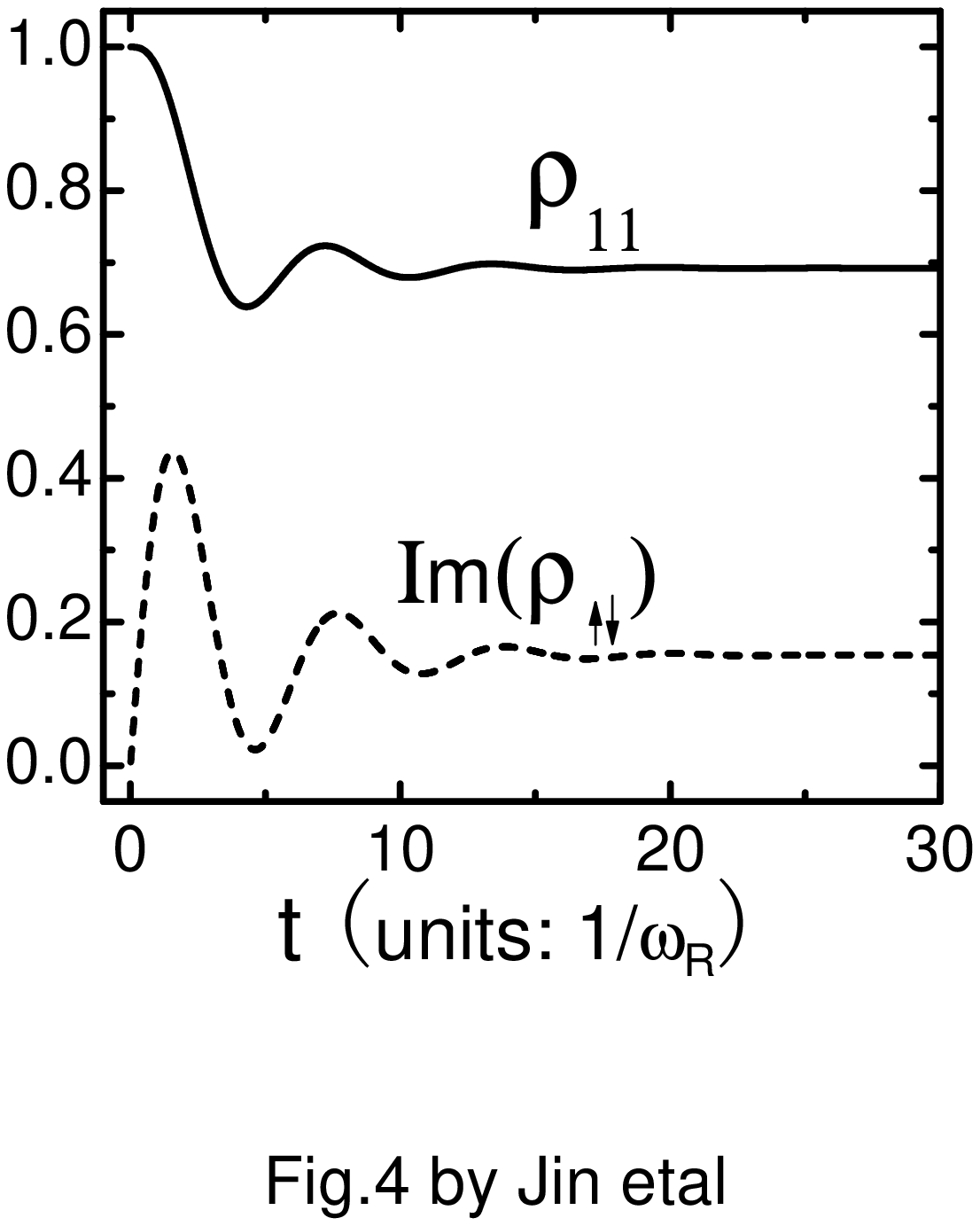}
\caption{ESR-induced coherent oscillation of the electron occupation
probability in the quantum dot (solid curve). The non-diagonal
element $\rho_{\uparrow\downarrow}$ describes the phase coherence of
the ESR (dashed curve), where the decay is caused by the
tunnel-coupling with the side reservoir (but not the back-action of
the QPC detector). The adopted parameters are the same as for Fig.\
3(b).  }
\end{figure}

The above behaviors of the noise spectrum are closely related to the
output current characteristics in different regimes,
as discussed previously in the context of Fig.\ 2.
Below we further discuss the {\it coherent} peaks in Fig.\ 3:
(i)
We would like to point out that the parameters $D$ and $D'$ are
related to the background noise $S_0=2eI_{\infty}$ (the first term of Eq. (9)),
meanwhile $D-D'$ would scale the magnitude of the excess noise which is caused
by the electron dynamics in the quantum dot (see the second term of Eq. (9)).
However, $D$ and $D'$ have no effect on the {\it intrinsic structure} of the noise spectrum.
(ii)
The height of the coherent peak in the noise spectrum is not very sensitive to
the tunneling rate $\Gamma$, but its evolution from ``birth" to ``death"
by increasing $\Gamma$ is evident.
Instead of its absolute value, the significant feature is the relative
magnitude of the central (zero frequency) peak and the side (coherent) peaks,
which essentially reflects the underlying dynamics, say,
being incoherent or partially coherent.
(iii)
The position of the coherent peak is around the Rabi frequency $\omega_R$.
This is understandable, because it is the Rabi oscillation
which causes the ``coherent" oscillating behavior of the occupation probability.
As we have pointed out, no quantum coherence is introduced and caused between the occupied
and unoccupied dot-states. However, the {\it coherent} flipping of the spin-up
and spin-down states of the electron inside the quantum dot would induce the
coherence-like oscillation of the occupation probability, with the same Rabi
frequency, as shown in Fig.\ 4.
The QPC can probe this short-time oscillating behavior and results in the
``coherent" peaks in the output noise spectrum.
Precisely speaking, the peak position does not locate exactly at the
Rabi frequency.
This can also be understood from the analytical result of the noise spectrum
in the measurement of solid-state charge qubit by the QPC \cite{Li05a}.
However, the slight shift of the peak position to higher frequency
with $\Gamma$ does not provide any interesting information.


In summary, we have presented a Monte Carlo simulation
for the single realization of continuous detection
of single electron spin.
The measurement scheme, say, based on the spin-to-charge conversion and
performing the electrical detection by a mesoscopic detector,
is of high interest for quantum computing.
The present study provides a theoretical description
for the measured RTS result \cite{Kou04,Mar04}, which may represent
an experimental breakthrough of single electron spin detection
in solid-states. The predicted feature in the output noise spectrum
reveals an interesting interplay of coherent oscillation and
incoherent tunneling. Its possible observation in experiment has
the value of directly illustrating the single electron quantum
dynamics in different regimes, i.e., from (partial) quantum to
classical.


\vspace{2ex} {\it Acknowledgments.} Support from the National
Natural Science Foundation of China and the Research Grants
Council of the Hong Kong Government are gratefully acknowledged.




\end{document}